\documentclass[sort&compress
    ,final            % use final for the camera ready runs 
    ,numberedheadings % uncomment this option for numbered sections
    ,cmfonts          % add further options here if necessary
  ]
  {aipproc}

\usepackage{subfigure}
\usepackage{amsmath,bm,graphicx}

\newcommand{\beq}{\begin{equation}}
\newcommand{\eeq}{\end{equation}} 
\newcommand{\beqa}{\begin{eqnarray}} 
\newcommand{\eeqa}{\end{eqnarray}}

 \def\bean{\begin{eqnarray*}}
 \def\eean{\end{eqnarray*}}
\newcommand{\ie}{{\it i.e.}}

\newcommand{\morder}[1]{{\cal O}\left(#1 \right)}
\newcommand{\eq}[1]{(\ref{#1})}

 \newcommand{\gsim}{\gtrsim}

\def\COMMENT#1{}

 \def\l{\left}
 \def\r{\right}

 \def\esim{\,\mathrel{\rlap{\lower0.2em\hbox{$-$}}\raise0.15em\hbox{\footnotesize $\hskip0.04em\sim$}}\,}
 \def\gsim{\mathrel{\rlap{\lower0.2em\hbox{$\sim$}}\raise0.2em\hbox{$>$}}}
 \def\ksim{\mathrel{\rlap{\lower0.2em\hbox{$\sim$}}\raise0.2em\hbox{$<$}}}

\newcommand{\be}{\begin{equation}}
\newcommand{\ee}{\end{equation}}
\newcommand{\bea}{\begin{eqnarray}}
\newcommand{\eea}{\end{eqnarray}}

\layoutstyle{6x9}

\begin{document}

\title{Collisional Energy Loss \\ of a Fast Parton in a QGP}

\author{St\'ephane Peign\'e}{
  address={SUBATECH, UMR 6457, Universit\'e de Nantes \\ Ecole des
Mines de Nantes, IN2P3/CNRS. \\ 4 rue Alfred Kastler, 44307 Nantes cedex 3, France}
}

\begin{abstract}

The suppression of hadron $p_T$ spectra in high energy central heavy-ion collisions compared
to proton-proton collisions, referred to as 'jet-quenching', is currently attributed to partonic energy loss 
in the hot medium created in the collision. The RHIC experiments show that at large enough $p_T$, 
hadron quenching is strong, and {\it of similar magnitude} for light and heavy flavours. This point
is difficult to understand in a parton energy loss scenario, where the energy loss is believed to 
be dominantly {\it radiative}, and quantitatively {\it different} for light partons and heavy quarks: 
gluon radiation off a heavy quark
is suppressed at small angles -- within the so-called {\it dead cone} -- and a heavy quark suffers less 
radiative energy loss than a light parton. 
In this context it is important to reconsider the {\it collisional} contribution to
partonic energy loss. Although it seems difficult to see how collisional losses could substantially 
increase heavy flavour quenching {\it without} simultaneously increasing light hadron quenching,
it is theoretically important to establish correct results for the heavy quark collisional energy loss. 

\end{abstract}

\maketitle

\renewcommand{\thefootnote}{\fnsymbol{footnote}}
\footnotetext{Talk presented at the Joint Meeting Heidelberg-Li\`ege-Paris-Wroclaw (HLPW08), 6-8 March, 2008, Spa, Belgium.}
\renewcommand{\thefootnote}{\arabic{footnote}}

\section{Jet-quenching phenomenology}
\label{sec1}

\begin{figure}[t]
\centering
\includegraphics[width=8cm]{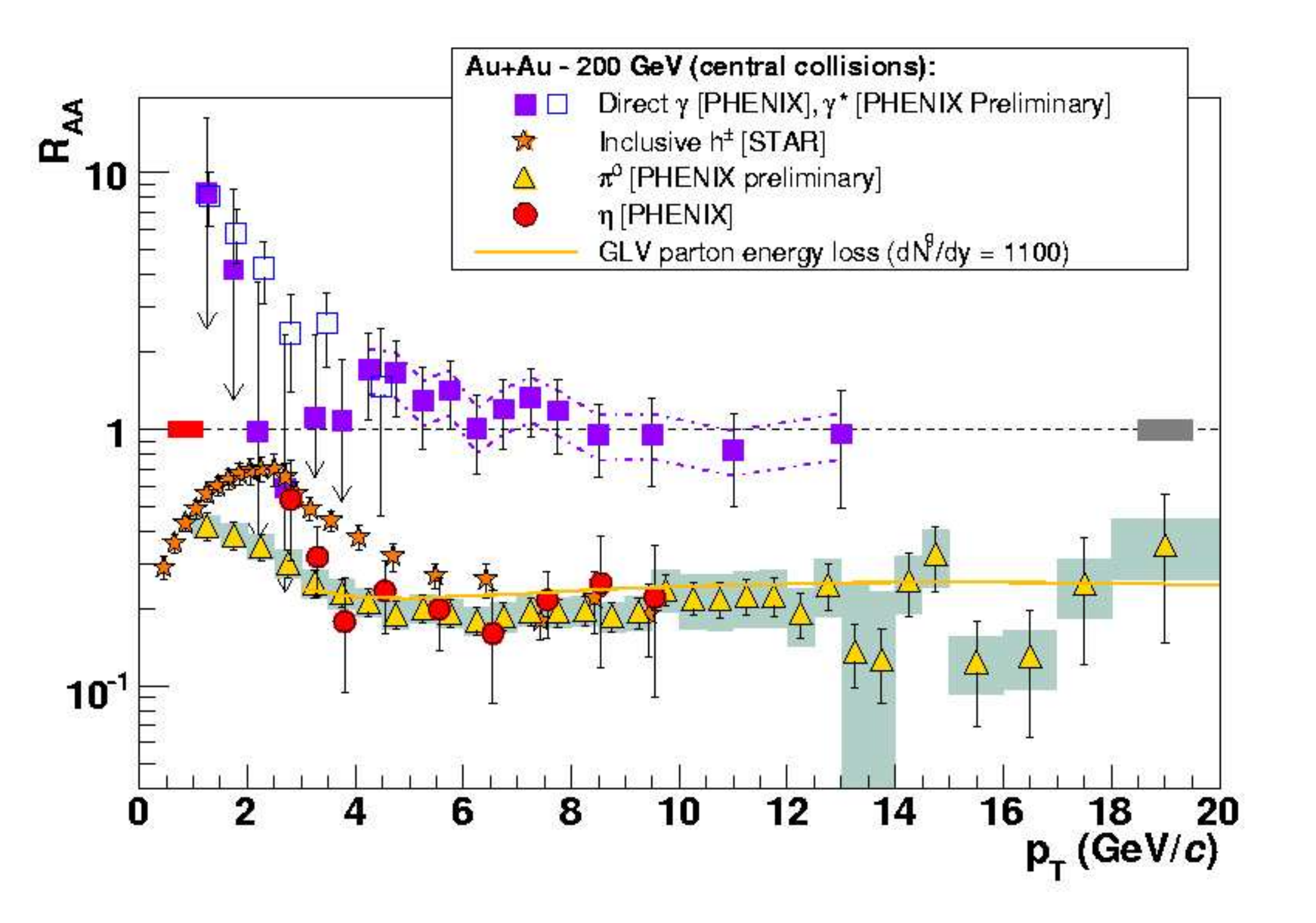}
\caption{Light hadron quenching. Direct photons are not quenched ($R_{AA}^{\gamma} \sim 1$), whereas light hadrons 
are ($R_{AA}^{h} \sim 0.2$). Figure taken from Ref.~\cite{d'Enterria:2006su}.}
\label{Fig:lighthadronquench}
\end{figure}

At the Relativistic Heavy Ion Collider (RHIC), the $p_T$ spectrum of {\it light} hadrons 
has been observed \cite{phenix,star} to be strongly suppressed in central ultrarelativistic heavy-ion collisions, as compared to 
proton-proton collisions. This is illustrated by the factor $R_{AA}^h(p_T)$,
\be
R_{AA}^h(p_T) = \frac{1}{N_{\rm {coll}}} \, 
{dN_{AA}^{h} \over d p_T } \biggr/  {dN_{p p}^{h} \over d p_T} \, ,
\ee
appropriately normalized by the number $N_{\rm {coll}}$ of binary nucleon-nucleon collisions in 
a nucleus-nucleus collision, to be much smaller than unity, as can be seen on Fig.~\ref{Fig:lighthadronquench}. 
For pions $R_{AA}^{\pi} \sim 0.2$ for $p_T > 4\,$GeV. 
%%%%%%%%%%%%%%%%%%%%%%%%%%%%%%%%%%%%%%%%%%%%%%%%%%%%%%%%%%%%%%%%%%%%%%%%%%%%%%%%%
%%%%%%%%%%%%%%%%%%%%%%%%%%%%%%%%%%%%%%%%%%%%%%%%%%%%%%%%%%%%%%%%%%%%%%%%%%%%%%%%%
This nuclear {\it attenuation} of hadron production, $R_{AA}^h <1$, has been quoted as {\it jet-quenching}, and 
is one of the most important signals for the new state of matter -- referred to below as 
a quark-gluon plasma or QGP -- produced in a (central) heavy-ion collision. 

The effect of jet-quenching was anticipated by Bjorken \cite{bj}, as a consequence of 
medium-induced partonic 
energy loss in the QGP. Although Bjorken illustrated his idea by performing an estimate
of the {\it collisional} loss of light partons, parton energy loss generally arises
from both collisional and {\it radiative} processes. In fact, for a parton energy $E \gg M, T$,
where $M$ is the parton mass and $T$ the plasma temperature, the radiative component 
is believed to be dominant. For a medium of large size we have \cite{Baier:1994bd}
\be
\Delta E_{rad}(L > L_{cr}) \sim \alpha_s^2 \sqrt{E T^3} \, L \gg \Delta E_{coll} \sim \alpha_s^2 T^2 \,L \, . 
\ee
The critical length $L_{cr}$ reads \cite{Baier:1994bd}
\be
L_{cr} \sim \frac{1}{\alpha_s T} \sqrt{\frac{E}{T}} \gg \lambda \sim \frac{1}{\alpha_s T}\, ,
\ee
where $\lambda$ is the parton's mean free path between two elastic scatterings. 
When $L$ is not large, $L< L_{cr}$, the radiative loss is modified \cite{Baier:1996kr,Zakharov:1997uu},
\be
\Delta E_{rad}(L < L_{cr}) \sim \alpha_s^3 T^3 L^2  \, ,
\ee
but still dominates over the collisional loss as long as $L > \lambda$. Thus it seems legitimate to neglect collisional 
energy loss in phenomenological studies of {\it light} hadron quenching, and to attribute hadron 
quenching solely to the {\it radiative} parton energy loss\footnote{For a light quark and for 
$L = 5\,$fm, a careful comparison of collisional and radiative
contributions within the same consistent model \cite{Zakharov:2007pj} gives the estimate 
$\Delta E_{coll}/\Delta E_{rad} \sim 0.25$ at $E \sim 20\,$GeV. 
Using $\lambda \simeq 1\,$fm, this is in agreement with our estimate $\Delta E_{coll}/\Delta E_{rad} \sim \lambda/L$
when  $L< L_{cr}$. In Ref.~\cite{Wicks:2005gt} a surprisingly large value $\Delta E_{coll}/\Delta E_{rad} \sim 0.75$ is 
used at $E \sim 20\,$GeV, obviously changing the qualitative understanding of quenching.}. 

\begin{figure}[t]
\centering
\includegraphics[width=7cm,clip=true]{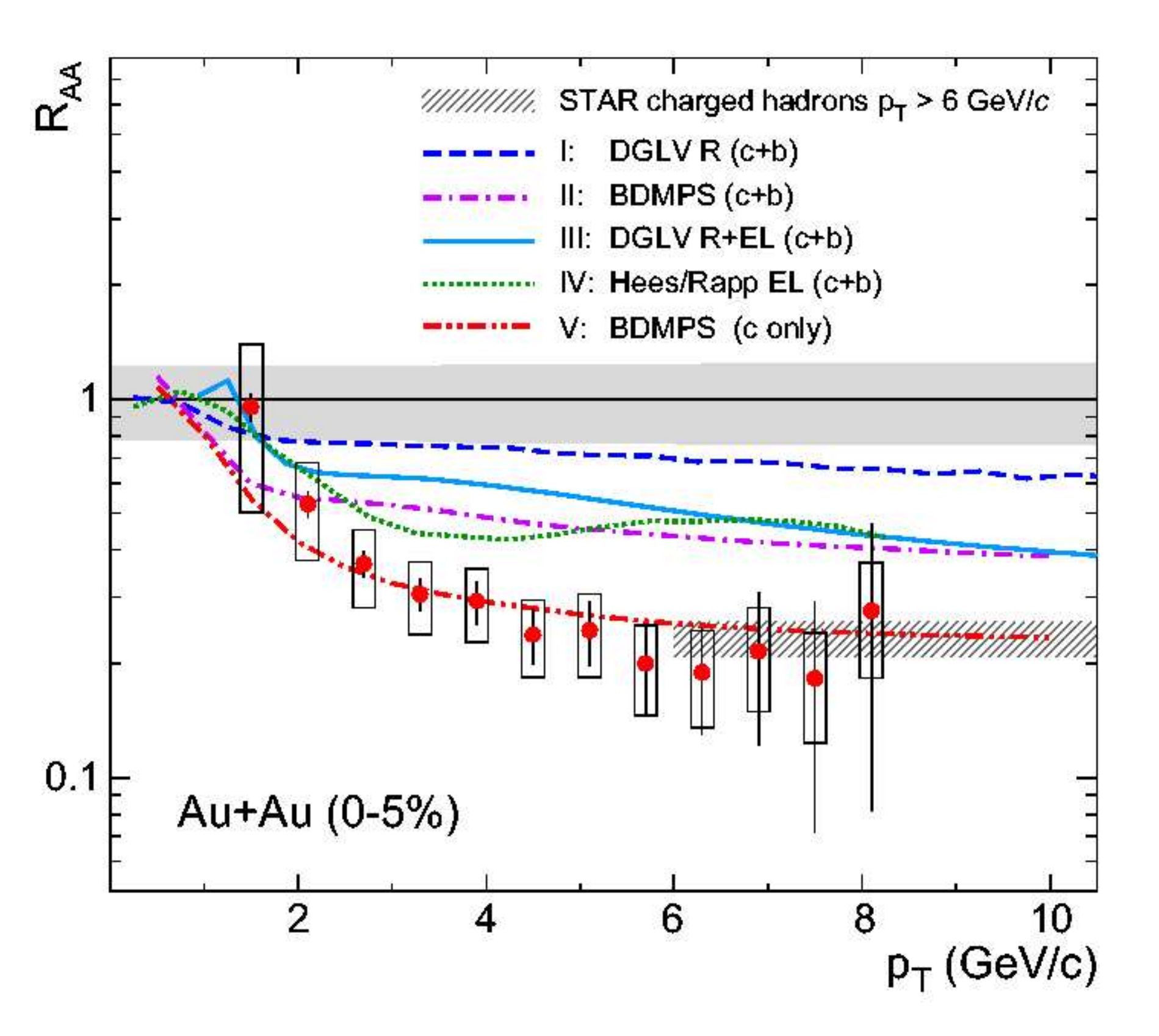}
\caption{The $R_{AA}$ factor for electrons from heavy 
D and B meson semi-leptonic decays. Figure taken from Ref.~\cite{Bielcik:2005wu}.}
\label{Fig:heavyflavourquench}
\end{figure}

The data on {\it heavy} flavour quenching \cite{Adler:2005xv,Abelev:2006db}, 
obtained indirectly by observing the electrons and positrons arising from D and B meson semi-leptonic 
decays, shows that the nuclear attenuation of heavy $c$ and $b$ quarks is as strong as that of 
light partons, $R_{AA}(Q) \simeq R_{AA}(q,g) \sim 0.2$ (see Fig.~\ref{Fig:heavyflavourquench}). 
%%%%%%%%%%%%%%%%%%%%%%%%%%%%%%%%%%%%%%%%%%%%%%%%%%%%%%%%%%%%%%%%%%%%%%%%%%%%%%%%%
%%%%%%%%%%%%%%%%%%%%%%%%%%%%%%%%%%%%%%%%%%%%%%%%%%%%%%%%%%%%%%%%%%%%%%%%%%%%%%%%%
This is difficult to explain when assuming that heavy quarks, similarly to light partons, 
lose energy dominantly through radiation. Indeed, for heavy quarks ($M \gg \Lambda_{\rm QCD}$), gluon radiation at an angle
$\theta_{rad} < M/p_T$ is suppressed \cite{Dokshitzer:2001zm}. This is the so-called `dead cone effect', leading
to a strong suppression (at not too large $p_T$), of the average radiative loss of a heavy quark compared to 
a light parton, 
\be
\Delta E_{rad}(b) < \Delta E_{rad}(c) < \Delta E_{rad}(q,g) \, .
\ee
If the electrons and positrons from heavy flavour decays observed in the experiment
would arise dominantly from D mesons, then the dead cone effect would not be too drastic, 
due the moderate value of the charm quark mass. In this case explaining the heavy flavour 
quenching with purely radiative $c$ quark energy loss might still be possible \cite{Zakharov:2007pj}.
However, the relative contributions from D and B decays are likely to be of the same order, as in  
$pp$ collisions \cite{bcratio}. The dead cone effect being stronger 
for $b$ than for $c$ quarks, this leads to a reduction of the effective radiative loss. As a consequence,
with similar contributions from $c$ and $b$ quarks, 
the current phenomenological models typically underestimate heavy flavour quenching.

Thus for {\it heavy} quarks, purely radiative energy loss seems insufficient to explain the observed attenuation. 
This has renewed the interest in the {\it collisional} part $\Delta E_{coll}$ 
of the parton energy loss. Could the collisional loss be much larger for a heavy quark than for 
a light parton, which would increase heavy flavour quenching without spoiling the reasonably good description of
light hadron quenching with purely radiative loss? 
 
We have reconsidered the {\it average} collisional loss of a heavy quark, and of 
a light parton. In the next section we summarize our calculations, which include the effect 
of the running coupling. We will see that the very definition 
of parton energy loss is different in the {\it tagged} heavy quark case and in the {\it untagged} light parton case.
As a result, in the high energy limit $E \gg M^2/T$, 
the heavy quark collisional loss contains an additional term -- a {\it collinear} logarithm 
$\propto \log{(ET/M^2)}$ arising from $u$-channel exchange -- which had been previously overlooked. 
This term makes the heavy quark collisional loss \eq{qcd-runningQ} 
slightly larger than the light quark loss \eq{qcd-runningq}, but is however small 
compared to the well-known {\it soft} logarithm $\propto \log{(ET/m_D^2)}$ ($m_D$ being the Debye mass of the
QCD plasma) arising from $t$-channel scattering. This is mainly due 
to a different associated color factor. We can thus approximate (see also \eq{Qqratio})
\be
E \gg M^2/T \Rightarrow \Delta E_{coll}(Q) \simeq  \Delta E_{coll}(q) \, .
\ee
We stress that the latter estimate holds in the ultra-high energy limit $E \gg M^2/T$. However, taking
$E \simeq p_T \simeq 20\,$GeV for the heavy quark energy, and $T  \simeq 200\,$MeV, we see that the data represented in 
Fig.~\ref{Fig:heavyflavourquench} correspond to the region $E \gg M^2/T$ for the charm quark, but not for the
bottom quark for which $E \ll M^2/T$. Since $\Delta E_{coll}(Q)$ is an increasing function of $E$ we must have 
\be
M_b^2/T \gg  E \gg M_c^2/T \Rightarrow \Delta E_{coll}(b) < \Delta E_{coll}(c) \simeq  \Delta E_{coll}(q) \, .
\ee

Thus in the kinematical region of Fig.~\ref{Fig:heavyflavourquench}, 
we expect both {\it collisional} and {\it radiative} contributions to the $b$ quark energy loss
to be suppressed compared to the $c$ or light quark case. In this respect the observed strong heavy flavour 
quenching, with an important contribution from $b$ quarks, remains a puzzle. 

This conclusion should however be considered as preliminary. Indeed, an {\it average} loss 
smaller for $b$ than for light quarks is a priori not incompatible with 
heavy and light flavour quenching being of similar magnitudes, due to the 
different energy loss {\it probability distributions} in both cases. 

\section{Parton collisional energy loss}

The collisional energy loss rate $dE/dx$ of an incoming parton of 4-momentum $P=(E,\vec{p})$
due to scattering off thermal particles of type $i$ ($i= q, g$) is given by
\be
\label{collrate}
\frac{dE}{dx} = \sum_{i} \frac{v^{-1}}{2E} \int_k\! \frac{{n_i(k)}}{2k} \int_{k'}\! \frac{{1 \pm n_i(k')}}{2k'} 
\int_{p'} \frac{(2\pi)^4}{2E'}\, \delta^{(4)}(P+K-P'-K')\, |{\cal M}_i |^2 \, \Delta E \, ,
\ee
where we denote $\int_k\ \equiv \int d^3\vec{k}/(2\pi)^3$. 
The expression \eq{collrate} represents the thermal average -- with appropriate Bose-Einstein or Fermi-Dirac thermal 
distributions -- of the elastic scattering amplitude squared $|{\cal M}_i |^2$, 
weighted by the energy loss in a single elastic scattering $\Delta E$ (to be defined precisely below). 
We will consider a fast parton of velocity
$v \to 1$. The appropriate sum and average of $|{\cal M}_i |^2$ over spin and color is 
implicit in \eq{collrate} and in the following. 

\subsection{Tagged heavy quark}

\begin{figure*}[t]
\centering
\begin{tabular}{c@{\hspace*{12mm}}c@{\hspace*{8mm}}c@{\hspace*{8mm}}c}
    \includegraphics[scale=0.7]{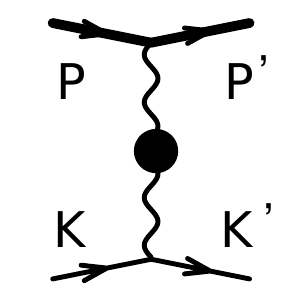}
   &
   \includegraphics[scale=0.7]{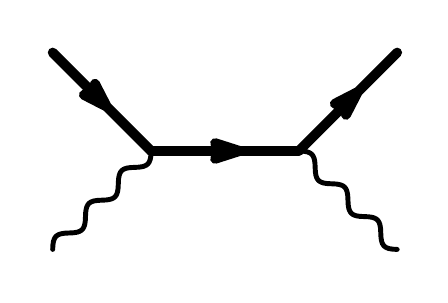}
   &
    \includegraphics[scale=0.7]{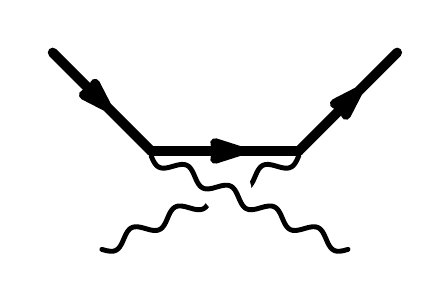}
   &
    \includegraphics[scale=0.7]{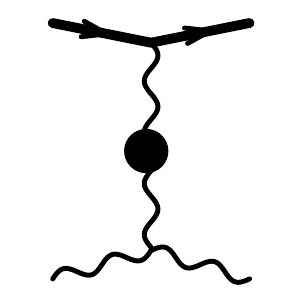}
   \\
     (a) & (b) & (c) & (d)
   \end{tabular}
   \caption{Amplitudes for heavy quark elastic scattering in a QGP. A curly line denotes a gluon.  
The blob in (a) and (d) denotes the resummed hard thermal loop gluon propagator, which is
necessary to screen the $t$-channel contribution in the infrared.}
  \label{fig:elasticamp}
  \end{figure*}

In the case the incoming particle is a heavy quark of mass $M \gg T$, we can neglect the presence of 
quarks and antiquarks of the same flavour in the plasma, and 
the scattering amplitude ${\cal M}_i$ is given by the diagrams shown in Fig.~\ref{fig:elasticamp}. 
In the following we will denote
$P'=(E',\vec{p}{\,'})$, $K=(k,\vec{k})$, $K'=(k',\vec{k}')$, and use the Mandelstam
invariants $s =(P+K)^2$, $u =(P-K')^2$, and $t =Q^2$, where $Q = K'-K = (\omega, \vec{q})$.
We will focus on the limit $E \gg M^2/T \gg M$, implying $s = M^2 + 2 P.K \sim \morder{E T} \gg M^2$.

%%%%%%%%%%%%%%%%%%%%%%%%%%%%%%%%%%%%%%%%%%%%%%%%%%%%%%%%%%%%%%%%%%%%%%%%%%%%%%%%%
%%%%%%%%%%%%%%%%%%%%%%%%%%%%%%%%%%%%%%%%%%%%%%%%%%%%%%%%%%%%%%%%%%%%%%%%%%%%%%%%%
When the heavy quark is {\it tagged} in the final state, the energy loss $\Delta E$ appearing in 
\eq{collrate} is simply the difference between the heavy quark initial and final energies, 
$\Delta E \equiv E- E' \equiv \omega$. We stress that for a tagged particle, events 
corresponding to {\it full stopping}, \ie, to $E' \ll E$ or equivalently 
to maximal energy transfer $\omega \simeq \omega_{\rm max} \simeq E$, contribute to $dE/dx$.

When $E \gg T$ the expression \eq{collrate} can be simplified to \cite{PPqed}
\beq 
\frac{dE}{dx}  =  \sum_{i} d_i \int_k n_i(k)
	\int_{|t|_{\rm min}}^{|t|_{\rm max}} d|t|\, \frac{d\sigma_i}{dt} \cdot \frac{|t|}{2k} \, ,
\label{eq:dEdx_asy}
\eeq
where $d_i$ is the degeneracy factor of target particles of type $i$, and 
\beq
\frac{d\sigma_i}{dt} = \frac1{16\pi (s -M^2)^2}\, \frac1{d\, d_i} \l| {\cal M}_i \r|^2 
\label{crossdef}
\eeq
is the corresponding differential cross section. In Eq.~\eq{eq:dEdx_asy} the bounds on $t$ are set by kinematics, 
$|t|_{\rm max} = (s -M^2)^2/s \simeq s$, and $|t|_{\rm min} = 0$. 

The result \eq{eq:dEdx_asy} is exact in the limit $E \gg T$ and can actually be simply understood. 
Noting that $|t| = 2 K.K' = 2 K.Q = 2 (k \omega - \vec{k}.\vec{q})$, the factor $\Delta E$ to be inserted in 
\eq{collrate} reads $\Delta E = \omega = |t|/(2k) + \vec{k}.\vec{q}/k$. From the result \eq{eq:dEdx_asy}, 
we see that effectively we can write $\omega = |t|/(2k)$, the term $\propto \vec{q}.\vec{k}$
vanishing after the angular integrals performed to obtain \eq{eq:dEdx_asy} starting from \eq{collrate}.
Without the last factor $\omega = |t|/(2k)$ in the r.h.s. of \eq{eq:dEdx_asy} we recognize 
$\rho \sigma \sim 1/\lambda$, with $\rho$
the spatial density of thermal scatterers, $\sigma$ the total elastic scattering cross section, 
and $\lambda$ the incoming particle's mean free path. Thus 
we can read the r.h.s. of \eq{eq:dEdx_asy} as $\sim \Delta E/\lambda \sim dE/dx$. 

In the following we will discuss, in a heuristic way, the leading logarithms contributing to 
$dE/dx$. We refer to Refs.~\cite{PPqed,PPqcd} for technical details.  

\vskip 5mm 
{\bf soft logarithm from $t$-channel exchange}
\vskip 5mm 

When $|t| \ll s$ the $t$-channel diagrams of Figs.~\ref{fig:elasticamp}a and \ref{fig:elasticamp}d contribute to 
the squared amplitude
as $\l| {\cal M}_{i} \r|^2 \propto \alpha_s^2 \, s^2/t^2$. Using \eq{crossdef}, we obtain from \eq{eq:dEdx_asy} 
the contribution
\be
\left. \frac{dE}{dx} \right|_{t\rm{-channel}}  \sim \alpha_s^2 T^2 \int_{|t|_{\rm min}}^{|t|_{\rm max}} \frac{d|t|}{|t|}
\sim \alpha_s^2 T^2 \left[ \ln\frac{E T}{m_D^2} + \morder{1} \right] .
\label{qed-t}
\ee
We used $|t|_{\rm max} \simeq s \sim ET$, which is justified to logarithmic accuracy, 
and $|t|_{\rm min} \sim m_D^2$ as an effective infrared cut-off, 
where $m_D^2 = 4 \pi \alpha_s T^2 (1 + n_f/6)$ is the Debye screening mass in the QGP \cite{BI}.
The logarithm in \eq{qed-t} thus arises from the broad logarithmic domain $m_D^2 \ll |t| \ll s$.

The inequality $|t| \ll s$ implies that the energy transfer in a single collision 
satisfies $\omega \ll E$. Thus the logarithm in \eq{qed-t} arises from 
kinematical configurations where the heavy quark is the {\it leading} parton in the 2-parton 
final state of the elastic scattering represented in Fig.~\ref{fig:elasticamp}. 

\vskip 5mm 
{\bf collinear logarithm from $u$-channel exchange}
\vskip 5mm 

The contribution from the square of the $u$-channel amplitude shown in Fig.~\ref{fig:elasticamp}c 
brings another logarithm in the heavy quark energy loss, 
arising from the domain $M^2 \ll |u| \ll s \sim ET$. Indeed, in this region we can approximate
$\l| {\cal M}_{i} \r|^2 \propto \alpha_s^2 \, s/|u|$, and $|t| \simeq s$ (recall that 
$s+u+t= 2M^2$). Changing variables from $t$ to $u$ in \eq{eq:dEdx_asy}, we thus obtain the contribution
\be
\left. \frac{dE}{dx} \right|_{u\rm{-channel}}  \sim \alpha_s^2 T^2 \int_{M^2}^{ET} d|u| \frac{1}{|u| s} |t| 
\sim \alpha_s^2 T^2 \left[ \ln\frac{E T}{M^2} + \morder{1} \right] .
\label{qed-u}
\ee

We emphasize that the logarithmic contribution \eq{qed-u} arises from $|t| \simeq s$, or
$\omega \simeq \omega_{\rm max} = E-M \simeq E$. Hence, in contrast to the soft logarithm of 
\eq{qed-t}, the collinear logarithm of \eq{qed-u} arises from full stopping of the incoming heavy 
quark, where the leading particle in the final state of Fig.~\ref{fig:elasticamp} is the struck thermal parton,
to which the incoming heavy quark has tranferred all its energy. 
This effect is the QCD analogue of QED Compton backscattering of laser beams, used to produce energetic
photons from energetic electrons. 
Full stopping must be taken into account in the case of a tagged particle, and this 
was overlooked in previous calculations of heavy quark energy loss \cite{TG,BTqcd}. 

Let us remark that the above energy loss process via ($u$-channel) Compton scattering 
is rare but very efficient. Neglecting logarithms and writing heuristically
\be
\left. \frac{dE}{dx} \right|_{u\rm{-channel}} \sim \frac{\langle \omega \rangle}{\lambda_{\rm Compton}} 
\sim \frac{E}{(E/\alpha_s^2 T^2)} \sim \alpha_s^2 T^2 \, ,
\ee
we see that despite a large Compton mean free path $\propto E$, the typical energy loss in Compton scattering
is large, $\langle \omega \rangle \sim E$, and the related rate $dE/dx$ can thus compete with the $t$-channel 
contribution.  

\vskip 5mm 
{\bf beyond leading logarithms}
\vskip 5mm 

In order to determine the constant beyond the leading logarithm in \eq{qed-t} and in \eq{qed-u}, the following 
procedure can be used \cite{PPqed,PPqcd}. Subtracting the 
leading logarithms $\int_{m_D^2}^{ET} d|t|/|t|$ and $\int_{M^2}^{ET} d|u|/|u|$ from the complete expression 
of $dE/dx$, the remaining integrals are dominated by either $|t| \sim m_D^2$ or $|t| \sim s \sim ET $.
An accurate calculation of the former integrals requires using the resummed hard thermal loop \cite{BI,HTL} 
gluon propagator in the $t$-channel, as pictured in Figs.~\ref{fig:elasticamp}a and \ref{fig:elasticamp}d.
The integrals dominated by $|t| \sim s \sim ET$ can be evaluated with bare exchanged propagators, but 
require using exact kinematics and exact expressions of the Born elastic scattering amplitude. 
Some previous attempts to calculate the heavy quark collisional loss 
beyond leading logarithm \cite{TG,BTqcd} were misleading,
for the simple reason that the collinear leading logarithm had been overlooked. 

Adding the leading logarithmic contributions \eq{qed-t}, \eq{qed-u}, and the constant 
$c(n_f) \simeq 0.146 \, n_f + 0.050$ determined in \cite{PPqcd} we obtain the heavy quark collisional loss
\beq
\frac{dE(Q)}{dx}  = \frac{4 \pi \alpha_s^2 T^2}{3} 
\left[ \left( 1 +\frac{n_f}{6} \right) \ln\frac{E T}{m_D^2} + \frac{2}{9} \ln\frac{E T}{M^2} 
+ c(n_f) \right] \, .
\label{qcd-tu}
\eeq
Note that the contribution from the collinear logarithm is reduced due to an associated 
smaller color factor \cite{PPqcd}.

\subsection{Untagged light parton}

Here we briefly discuss the case of a {\it light} incoming parton. 
When there is no heavy-quark tagged jet, it is impossible to know whether the detected jet
arises, in the elastic scattering, from the scattered incoming parton of from the struck thermal parton.
Moreover, a light incoming parton can be annihilated, due to the presence 
of partons of the same flavour in the plasma. Thus, in the {\it untagged} case, the loss $\Delta E$ in 
elastic scattering must be defined with respect to the {\it leading} parton in the final state.
Defining momenta as in Fig.~\ref{fig:elasticamp} we have (neglecting $k \sim T$ compared to $E$)
\be
\label{untaggedloss}
\begin{array}{ccc}
\omega < E/2 & \Rightarrow & \Delta E  = P_0 - P_0' = E- E' = \omega  \\ \\ 
\omega > E/2 & \Rightarrow & \Delta E  = P_0 - K_0' \simeq E - \omega < E/2 
\end{array} 
\ee
Since $\Delta E < E/2$, there is no full stopping with this definition. 

Thus when $\omega > E/2$, the {\it apparent} energy loss, defined with respect to the leading parton,
is not $\omega$ as in the heavy-quark tagged case, but $E - \omega$. Let us express the 
condition $\omega > E/2$ using Mandelstam invariants. We have
\be
\label{twoeq} 
\begin{array}{cc} 
\ & |t| = 2 K.K' = 2k (k+\omega) (1- \cos{\theta_{kk'}})  \\ \\ 
\ & s \simeq  2 P.K \simeq 2 E k (1-\cos{\theta_{kp}}) 
\end{array} 
\ee
where we used $s \gg M^2$ and $E \gg M$ in the second line. When $\omega > E/2 \gg k \sim T$, 
it is easy to check from energy-momentum conservation that $\vec{p}$ and $\vec{k}'$ are quasi-collinear, 
implying $\theta_{kk'} \simeq \theta_{kp}$. Taking the ratio of the two equations in \eq{twoeq} we thus obtain
\be
\omega > E/2 \Rightarrow \frac{|t|}{s} \simeq \frac{\omega}{E} \Rightarrow  \frac{|u|}{s} \simeq \frac{E-\omega}{E} \, .
\ee
To get the last equality we used $s+u+t= 2M^2$ and assumed that $|u|/s \sim \morder{1}$, which will be justified in the
following. Reexpressing \eq{untaggedloss} in terms of Mandelstam invariants we get 
\be
\label{untaggedlossinv}
\begin{array}{ccc}
|t| < s/2 & \Rightarrow & \Delta E /E  \simeq  |t|/s   \\ \\ 
|t| > s/2 & \Rightarrow & \Delta E /E  \simeq  |u|/s  
\end{array} 
\ee

When $|t| > s/2$, the energy loss weight in \eq{collrate} is simply obtained by replacing 
$|t| \to |u|$ in the integrand. 
In the untagged case, the expression \eq{eq:dEdx_asy} of the energy loss is thus modified to 
\beq 
\left. \frac{dE}{dx} \right|_{\rm{untagged}}  =  \sum_i d_i \int_k n_i(k)
\left[	\int_{|t|_{\rm min}}^{s/2} d|t|\, \frac{d\sigma_i}{dt} \cdot \frac{|t|}{2k}
+ \int_{s/2}^{|t|_{\rm max}} d|t|\, \frac{d\sigma_i}{dt} \cdot \frac{|u|}{2k} \right] \, .
\label{eq:dEdx_asy-untagged}
\eeq
We can readily see that the first term of \eq{eq:dEdx_asy-untagged} will yield the same $t$-channel leading logarithm 
as in the tagged case, but that the second term does not produce any $u$-channel logarithm, due to the 
energy loss weight $\propto |u|$ instead of $\propto |t|$. 
The term of $\l| {\cal M}_i \r|^2$ which is $\propto s/|u|$ indeed contributes to the second term of 
\eq{eq:dEdx_asy-untagged} as 
\be
\int_k \frac{n_i(k)}{2k} \int^{s/2} d|u|\, \frac{\alpha_s^2}{s |u|} 
\cdot |u| \sim \alpha_s^2 T^2 \, , 
\label{subleading}
\ee
where the integral over $u$ is dominated by $|u| \sim \morder{s} \sim \morder{ET}$. The contribution \eq{subleading} 
is subleading compared to the $t$-channel logarithmic contribution.

As a result, the energy loss of an untagged parton, to logarithmic accuracy, is given by the first term 
of \eq{qcd-tu} only,
\beq
\left. \frac{dE}{dx} \right|_{\rm{untagged}} = C_R \pi \alpha_s^2 T^2 
\left[ \left( 1 +\frac{n_f}{6} \right) \ln\frac{E T}{m_D^2} + \morder{1} \right] \, ,
\label{qcd-lightquark}
\eeq
where $C_R = C_F = 4/3$ for a quark and $C_R = C_A = 3$ for a gluon.
The expression \eq{qcd-lightquark} is Bjorken's result \cite{bj} and is specific to the case
of an {\it untagged} parton. We stress the importance of defining correctly the {\it observable} (tagged or 
untagged) parton energy loss in order to obtain meaningful results, even to leading logarithmic accuracy.

\subsection{Running coupling} 

The results \eq{qcd-tu} and \eq{qcd-lightquark} obtained in the fixed coupling approximation 
are not truly predictive, since the scale at which to evaluate $\alpha_s$ is not specified.
In order to obtain predictive results, we must repeat our calculations including the effect
of running coupling. 

In the case of $t$-channel scattering (Figs.~\ref{fig:elasticamp}a and \ref{fig:elasticamp}d), we must 
set the scale of $\alpha_s$ to $\sim \morder{|t|}$, as dictated by renormalization in PQCD
at zero temperature \cite{APrunning}. Setting this scale in the factor $\alpha_s^2$ of the $2 \to 2$ cross section 
amounts to perform a specific 
all-order resummation, corresponding physically to the final partons being accompanied by collinear gluon radiation. 
The true {\it exclusive} $2 \to 2$ cross section involving charged (coloured) particles vanishes \cite{Dokshitzer:1991wu}.

With running coupling, the $t$-channel leading logarithm in \eq{qed-t} becomes
\beq
 \alpha_s^2 \int^{ET}_{m_D^2} \frac{d|t|}{|t|} 
\longrightarrow 
\int^{ET}_{m_D^2} \frac{d|t|}{|t|} \alpha_s^2(|t|) \, .
\label{rule1}
\eeq 
Using $\alpha_s(|t|) = \left[ 4\pi \beta_0 \ln{(|t|/\Lambda^2)}\right]^{-1}$, 
where $\beta_0 = (11-\frac{2}{3} n_f)/(4\pi)^2$, we find that using a running coupling amounts to the replacement
\beq
\alpha_s^2 \ln{(ET/m_D^2)}
\,\longrightarrow\,  \alpha_s(m_D^2) \alpha_s(ET) \ln{(ET/m_D^2)}  \, .
\label{rule2}
\eeq
As noted in \cite{APrunning}, the latter result remains finite when $E \to \infty$. 
The asymptotic collisional loss is bounded, instead of logarithmically enhanced
as suggested by the fixed coupling result \eq{qcd-tu}. 

For the $u$-channel contribution (Fig.~\ref{fig:elasticamp}c), $\alpha_s$ should be evaluated at a scale
$\sim \morder{|u|}$. Similarly to the preceding discussion, the $u$-channel collinear logarithm appearing in \eq{qed-u} 
is modified as
\beq
\alpha_s^2 \ln{(ET/M^2)}
\longrightarrow  \alpha_s(M^2) \alpha_s(ET) \ln{(ET/M^2)} \, .
\label{rule3}
\eeq

We have already mentioned that the constant appearing next to the leading logarithms 
(in \eq{qcd-tu} or \eq{qcd-lightquark}) is given 
by integrals where the typical value of $t$ is determined by either $|t| \sim m_D^2$
or $|t| \sim ET$. Since there is no logarithmic spread there, the terms $\propto \alpha_s^2$ in \eq{qcd-tu} and \eq{qcd-lightquark}
which are not logarithmically enhanced can be evaluated at a scale chosen arbitrarily between $m_D^2$ and $ET$.

Using \eq{qcd-tu}, \eq{qcd-lightquark}, and the modifications \eq{rule2}, \eq{rule3} required by running coupling,
we can write the collisional loss of a fast ($E \gg M^2/T \gg M$) heavy quark as 
\beq
\frac{dE(Q)}{dx}  = \frac{4}{3} \pi  \alpha_s(m_D^2) \alpha_s(ET) T^2
\left[ \left( 1 +\frac{n_f}{6} \right)  \ln\frac{E T}{m_D^2} 
+ \frac{2}{9} \frac{\alpha_s(M^2)}{\alpha_s(m_D^2)} \ln\frac{E T}{M^2} 
+  c(n_f) \right] \, ,
\label{qcd-runningQ}
\eeq
and the collisional loss of a light energetic ($E \gg T$) parton as 
\beq
\left. \frac{dE}{dx} \right|_{\rm{untagged}}  = C_R \pi \alpha_s(m_D^2) \alpha_s(ET) T^2
\left[ \left( 1 +\frac{n_f}{6} \right)  \ln\frac{E T}{m_D^2}  + \morder{1}\right] \, .
\label{qcd-runningq} 
\eeq
We refer to \cite{PPqcd} for a discussion of the theoretical uncertainty.

Due to the additional logarithm in \eq{qcd-runningQ}, when $E \gg M^2/T$ the collisional loss 
is (slightly) larger for a heavy than for a light (untagged) quark. Taking $E \to \infty$ and $n_f =3$, 
\be
\label{Qqratio}
\frac{dE(Q)/dx}{dE(q)/dx}  
%% \left. \frac{dE}{dx} \right|_{Q}  \biggr/  \left. \frac{dE}{dx} \right|_{q} 
\mathop{\ \ \longrightarrow\ \ }_{E \to \infty} 1 + \frac{4}{27} \frac{\alpha_s(M^2)}{\alpha_s(m_D^2)} 
< 1 + \frac{4}{27} \simeq 1.15 \, .
\ee

\section{Summary}

We discussed the average collisional loss $dE/dx$ of a heavy quark crossing a QGP, in the limit $E \gg M^2/T$,
and briefly reviewed the case of an energetic $E \gg T$ light parton.
For fixed $\alpha_s$, at leading order $dE/dx \propto \alpha_s^2$, 
with a coefficient which is logarithmically enhanced. A soft logarithm 
is present for both heavy and light partons. A collinear logarithm contributes additionally 
to the heavy quark loss, but {\it not} to the light parton loss, due the difference between the definition of energy loss 
for a {\it tagged} heavy quark and for an {\it untagged} light parton. Implementing the running of 
$\alpha_s$ we obtained the corresponding results \eq{qcd-runningQ} and \eq{qcd-runningq}.

At very high energy the heavy quark collisional loss is slightly enhanced compared to the
light quark loss, see \eq{Qqratio}. However, as mentioned in section \ref{sec1}, 
for $E \sim p_T \sim \morder{20\,\rm{GeV}} \ll M_b^2/T$, the $b$ quark collisional (as well as radiative) loss 
should be reduced compared to the light quark case, challenging our understanding of 
heavy flavour quenching.

\end{document}